\documentclass{ptapap}
\usepackage{url}

\author{Gianfranco De Zotti}[OaPd]
\affil[OaPd]{INAF-Osservatorio Astronomico di Padova, \\ Vicolo dell'Osservatorio 5, I-35122 Padova, Italy
}

\title{Prospects for next generation Cosmic Microwave Background experiments}

\begin{document}

\maketitle

\def\simlt{\mathrel{\rlap{\lower 3pt\hbox{$\sim$}}\raise 2.0pt\hbox{$<$}}}
\def\simgt{\mathrel{\rlap{\lower 3pt\hbox{$\sim$}} \raise
2.0pt\hbox{$>$}}}
\def\lsim{\,\lower2truept\hbox{${<\atop\hbox{\raise4truept\hbox{$\sim$}}}$}\,}
\def\gsim{\,\lower2truept\hbox{${> \atop\hbox{\raise4truept\hbox{$\sim$}}}$}\,}

\begin{abstract}
In this lecture, after a synthetic review of measurements of CMB temperature
anisotropies and of their cosmological implications, the theoretical
background of CMB polarization is summarized and the concepts of the main
experiments that are ongoing or are being planned are briefly described.

\end{abstract}

\section{Introduction}

Although Cosmic Microwave Background (CMB) experiments, and notably the highly
successful WMAP and \textit{Planck} space missions, have put on a solid basis
the so-called standard cosmological model and have yielded very accurate
determinations of its basic parameters, the information content of the CMB has
not been fully exploited yet. In particular, \textit{Planck}  can be considered
the definitive mission about CMB temperature anisotropies on scales $\ge
5\,$arcmin. However, the sensitivity of \textit{Planck} to CMB polarization was
not sufficient to extract all the information carried by it. To get such
information we need polarization measurements one or more orders of magnitude
better than \textit{Planck}. The present Holy Grail is the detection of
$B$-mode polarization that would provide a direct test of the notion of
cosmological inflation, which is at the basis of the overwhelming majority of
current models of the early Universe. The importance of this scientific goal
has triggered a lot of ground-based, sub-orbital and space-borne projects.

The plan of this lecture is the following. Section~\ref{sect:temp} contains a
synthetic review of the accurate measurements of CMB temperature anisotropies
provided by the \textit{Planck} satellite and of their cosmological
implications. Section~\ref{sect:pol} deals with CMB polarization anisotropies.
Section~\ref{sect:next} gives a short description of the main ongoing or
planned next generation CMB experiments. Finally, Sect.~\ref{sect:conclusions}
summarizes the main conclusions.

\section{CMB temperature anisotropies }\label{sect:temp}

\subsection{Where do we stand?}

Starting with the COBE detection of the CMB anisotropy in \citet{Smoot1992},
the mapping of the primordial CMB anisotropies in temperature and polarization,
produced by the WMAP and, with higher precision, by the \textit{Planck}
satellite has allowed us to characterize the initial cosmological perturbations
at about the percent level.

Anisotropies are more conveniently studied in terms of multipoles, i.e. in the
Legendre (for a sphere) or Fourier (on a planar surface) space. \textit{Planck}
has determined the CMB temperature power spectrum down to fundamental limits up
to multipoles $\ell \sim 1500$ i.e. down to $\sim 5\,$arcmin angular scales
\citep{PlanckCollaborationXI2016}. The temperature power spectrum determination
has been extended up to $\ell \sim 3000$ by higher resolution ground based
experiments such as the South Pole Telescope \citep[SPT;][]{George2015,
Henning2017} and the Atacama Cosmology Telescope \citep[ACT;][]{Crites2015,
Louis2017}.

Remarkably, the data can be accounted for by the standard
\textbf{six-parameter} cold dark matter model with a cosmological constant
($\Lambda$CDM model) and a power spectrum of primordial density perturbations
having a simple power-law form:
\begin{equation}
\mathcal{P}(k)= A_s\left(\frac{k}{k_\star}\right)^{n_s-1},
\end{equation}
where $k_\star$ is a pivot scale, generally fixed to $0.05\,\hbox{Mpc}^{-1}$.

In addition to the 2 parameters characterizing the perturbation spectrum, the
model includes 4 non-inflationary parameters: the Hubble constant $H_0$; the
baryon density $\omega_b=h^2 \Omega_b$ (where
$h=H_0/100\,\hbox{km}\,\hbox{s}^{-1}\,\hbox{Mpc}^{-1}$ and $\Omega_b$ is the
baryon density in units of the critical density), $\omega_c=h^2 \Omega_c$
($\Omega_c$ being the dark matter density in units of the critical density);
the optical depth due to re-ionization, $\tau$.

With these 6 parameters a very good fit of the temperature power spectrum is
achieved. There is no statistically significant evidence compelling us to add
more parameters, despite the many extensions that have been explored
\citep{PlanckCollaborationXVI2014, PlanckCollaborationXXII2014,
PlanckCollaborationXIII2016, PlanckCollaborationXX2016}.

Importantly, the model includes only the adiabatic growing mode for primordial
fluctuations, as predicted for inflation driven by a single scalar field. No
statistically significant evidence was uncovered showing that isocurvature
modes were excited \citep{PlanckCollaborationXXII2014,
PlanckCollaborationXX2016}, which is possible in multi-field inflationary
models.

One of the most significant findings, first made by WMAP at modest statistical
significance and later by \textit{Planck} at much higher significance
\citep{Bennett2013}, was that the primordial power spectrum is not exactly
scale invariant: in other words, $n_s\neq 1$
\citep{PlanckCollaborationXXII2014, PlanckCollaborationXX2016}. A value of
$n_s$ very close to unity would require an unnatural fine tuning.

\textit{Planck} also set tight constraints on deviations of the statistics of
primordial fluctuations from a Gaussian distribution
\citep{PlanckCollaborationXXIV2014, PlanckCollaborationXVII2016}. These bounds
rule out at high statistical significance many non-standard inflationary models
predicting a level of non-Gaussianity allowed by WMAP \citep{Bennett2013}.

\subsection{Is there room to improve over \textit{Planck}?}

To answer this question let us consider the errors on the CMB power spectra,
$C_\ell$. If the underlying stochastic process is nearly Gaussian, as found by
\textit{Planck}, we have, approximately \citep{Knox1995}
\begin{equation}
\left({\Delta C_\ell \over C_\ell}\right)_{\rm rms} = \sqrt{{2\over f_{\rm sky}(2\ell+1)}}\frac{C_\ell + N_\ell}{C_\ell},
\end{equation}
where $f_{\rm sky}=A/4\pi$ is the sky fraction surveyed ($A$ is the solid angle
covered by the survey), $C_\ell$ is the expected or theoretical power spectrum
and $N_\ell$ is the power spectrum of the measurement noise.

In the ideal case of negligible noise ($C_\ell \gg N_\ell$), the uncertainty is
due to the ``cosmic variance''\footnote{It should be noted that the cosmic
variance refers to $f_{\rm sky}=1$. If $f_{\rm sky} < 1$ we are dealing with
the ``sample variance'' that is different for different surveyed areas.}: we
cannot further improve our determination of the power spectrum because it is a
statistics of a stochastic process and we have only one realization of the sky.
Thus, with regard to the determination of the power spectrum, a further
decrease of $N_\ell$ is of only marginal added value.

At high multipoles (small angular scales) the cosmic variance decreases and
eventually the error becomes noise-dominated. The noise power spectrum is
boosted as we reach the multipole corresponding to the angular resolution,
$\theta$, of the instrument: $\ell \simeq 60^\circ/\theta$, with $\theta$ in
degrees \citep{White1994}. If the beam profile is Gaussian with full width at
half maximum (FWHM) $\theta$, the noise is boosted exponentially:
\begin{equation}\label{eq:boost}
N_{\ell, \rm boost}\simeq N_\ell \exp\left(\frac{\ell^2\theta^2}{2\sqrt{2\ln2}}\right).
\end{equation}
\textit{Planck} sensitivity was so high that it was at the cosmic variance
limit up to $\ell \simeq 1500$, i.e. up to its maximum resolution. At higher
multipoles its noise blows up according to eq.~(\ref{eq:boost}). A higher
resolution ($\theta \sim 2'$) mission, with the high sensitivity allowed by
modern technology, could reach the cosmic variance limit up to $\ell \simeq
3000$ \citep[cf. the left panel of Fig.~1 of][]{Finelli2016}.

Note that also ground-based experiments can reach the cosmic variance limit at
high-$\ell$'s. On the other hand, at low-$\ell$'s a full-sky coverage, possible
only from space, is necessary.

Note also that the high-$\ell$ regime corresponds to the damping tail of CMB
an\-iso\-tro\-pies. The CMB signal is weak and its recovery requires a very
accurate removal of foreground emission, dominated by point sources. On the
other hand, high resolution surveys have a great potential to enhance our
knowledge of extragalactic sources \citep{DeZotti2015, DeZotti2016}.

There is much more room for improvement on polarization measurements. Here the
\textit{Planck} noise is well above the cosmic variance limit ($f_{\rm sky}=1$)
for the $EE$ power spectrum (defined in Sect.~\ref{sect:pol})  except for the
lowest multipoles \citep[cf. the right-hand panel of Fig.~1 of][]{Finelli2016}.

As illustrated in that figure, the higher sensitivity (compared to
\textit{Planck}) provided by the modern technologies  allows a substantial
improvement, even without a higher resolution.

\section{CMB polarization anisotropies}\label{sect:pol}

\subsection{The quest for information on the physics of primordial inflation}

At the sensitivity level of presently available data, the initial conditions of
the universe are described by just two numbers: the amplitude of primordial
curvature perturbations, $A_s$, and its spectral index $n_s$. But at this level
the form of the power spectrum follows from the weakly broken scaling symmetry
of the inflationary space-time and is therefore rather generic.

Only upper limits are available on other quantities that can provide detailed
information about the microphysical origin of inflation such as \citep[cf.
Table~1 of][]{Finelli2016}: the ``running'' of the spectral index of scalar
perturbations i.e. its dependence on the scale of perturbations, ${\rm
d}n_s/{\rm d}\ln k$; the amplitude, $A_t$, and spectral index, $n_t$, of tensor
perturbations; the tensor to scalar ratio $r=A_t/A_s$;  the spatial curvature,
$\Omega_k$; the non-Gaussianity of primordial perturbations; the amount of
isocurvature perturbations, yielded by extra fields at inflation; topological
defects, whose detection would be informative on the end of inflation.

With future, much higher sensitivity information,  we expect much more detailed
information about the physics of the inflationary era. In particular, crucial
information is expected from primordial $B$-mode polarization, produced by
tensor perturbations.

\subsection{Physics of CMB polarization}

The differential cross-section for Thomson scattering into the solid angle
element ${\rm d}\Omega$ of unpolarized radiation is \citep{RybickiLightman1979}
\begin{equation}
\frac{{\rm d}\sigma}{{\rm d}\Omega}=\frac{1}{2} r_0^2(1+\cos^2\theta),
\end{equation}
where $r_0=e^2/m_e c^2 \simeq 2.82\times 10^{-13}\,$cm  is the ``classical
electron radius''  and $\theta$ is the angle between the scattered and the
incident photon. The incident light sets up oscillations of the target electron
in the direction of the electric field vector $E$, i.e. the polarization. The
scattered radiation intensity thus peaks in the direction normal to, with
polarization parallel to, the incident polarization.

In the case of unpolarized isotropic radiation, the polarization induced by
scattering in two perpendicular directions balance and the radiation remains
unpolarized. The same happens in the case of a dipole anisotropy. But if the
radiation possesses a quadrupole anisotropy the scattered radiation acquires a
linear polarization. A reversal in sign of the temperature fluctuation
corresponds to a $90^\circ$ rotation of the polarization, which reflects the
spin-2 nature of polarization \citep{HuWhite1997}.

\subsection{$E$ and $B$ modes}

Linear polarization is measured by the Stokes parameters $Q$ and $U$. However,
these quantities depend on an arbitrary choice of the coordinates. Under a
coordinate rotation by an angle $\phi$ they transform as:
\begin{equation}
\left(\begin{array}{c} \tilde{Q} \\ \tilde{U} \end{array} \right) =
\left(\begin{array}{cc} \cos(2\phi)\ \  \sin(2\phi) \\\!\! -\sin(2\phi)\ \  \cos(2\phi) \end{array} \right)
\left(\begin{array}{c} {Q} \\ {U} \end{array} \right).
\end{equation}
This dependence on coordinates is not very convenient. To define a
coordinate-independent quantity for the distribution of polarization patterns
in the sky we need to go to Fourier space. This allows us to describe the
polarization pattern by its orientation relative to itself.

There are two directions picked out by a polarization pattern: that which is
picked out by its orientation and that which is picked out by its amplitude.
The amplitudes of the polarization patterns  are modulated in space by the
plane wave they are sitting on.

We can then construct two quantities, $E$ and $B$ such that \citep{Seljak1997,
ZaldarriagaSeljak1997, Kamionkowski1997}:
\begin{eqnarray}
Q(\theta) &=& \int \frac{d^2\ell}{(2\pi)^2}\left(E_\ell\cos(2\phi_\ell)- B_\ell\sin(2\phi_\ell)\right) \exp(i\ell\theta) \nonumber \\
U(\theta) &=& \int \frac{d^2\ell}{(2\pi)^2}\left(E_\ell\sin(2\phi_\ell)+ B_\ell\cos(2\phi_\ell)\right) \exp(i\ell\theta). \nonumber
\end{eqnarray}
Following E.~Komatsu\footnote{Lectures given at the XIII School of Cosmology
November 12--18, 2017, Carg\`ese, available at
\url{http://www.cpt.univ-mrs.fr/~cosmo/EC2017/Programme17_a.html}} let us
consider $Q$ and $U$ that are produced by a single Fourier mode. Taking the
x-axis to be the direction of a wavevector, we obtain:
\begin{eqnarray}
Q(\theta) &=& E_\ell \exp(i\ell\theta) \nonumber \\
U(\theta) &=& B_\ell \exp(i\ell\theta). \nonumber
\end{eqnarray}
Thus the $E$-mode is the Stokes $Q$, defined with respect to the wavevector as
the x-axis; the $B$-mode is the Stokes $U$, defined with respect to the
wavevector as the y-axis. The $E$-mode describes the polarization directions
parallel or perpendicular to the wavevector; the $B$-mode describes the
polarization directions tilted by $45^\circ$ with respect to the wavevector.
These definitions no longer depend on an arbitrary choice of the coordinates.

The $E$-mode is unchanged under a reflection (even parity), while the $B$-mode
changes sign (odd parity). This implies that the $EB$ and $TB$ cross-power
spectra vanish for parity-preserving fluctuations because $EB$ and $TB$ change
sign under parity flip. In addition to the 3 auto-power spectra ($TT$, $EE$ and
$BB$) there is only one non vanishing temperature-polarization cross-spectrum,
namely $TE$.

Density perturbations just generate parallel polarization and so generate only
$E$-mode polarization. Gravitational waves generate both and so have a
component of $B$-mode polarization\footnote{
\url{http://background.uchicago.edu/~whu/intermediate/Polarization/polar5.html}.}.

As the CMB radiation possesses a primary quadrupole moment, Thomson scattering
between the CMB photons and free electrons generates linear polarization. This
is the case both at recombination and at re-ionization. Re-scattering of the
CMB photons at reionization generates a new polarization anisotropy on larger
angular scales than at recombination because the horizon has grown to a much
larger size by that epoch.

\subsection{Measurements of CMB polarization and re-ionization}

Measurements of the $EE$ and $TE$ power spectra have been obtained by several
experiments: WMAP \citep{Page2007, Nolta2009, Larson2011, Bennett2013}, SPTpol
\citep{George2015, Henning2017}; ACTpol \citep[ACT;][]{Crites2015, Louis2017};
POLARBEAR \citep{PolarbearCollaboration2014, PolarbearCollaboration2017};
BICEP/Keck Array \citep{Barkats2014}, among others. The most accurate
determinations have been provided by the \textit{Planck} mission
\citep{PlanckCollaborationXI2016}.

A very important information provided by these power spectra is about the
cosmic re-ionization. The location of the anisotropy peak in the CMB power
spectrum relates to the horizon size at the new ‘last scattering’ and thus
depends on the ionization redshift $z_{\rm reion}$. A fitting formula was given
by \citet{Liu2001}:
\begin{equation}
\ell_{\rm peak} \simeq 0.74(1+ z_{\rm reion})^{0.73}\Omega_m^{0.11},
\end{equation}
where $\Omega_m$ is the matter (baryons$+$dark matter) density in units of the
critical density. The  peak amplitude is a measure of the optical depth to
reionization, $\tau$.

The re-ionization produces a low-$\ell$ peak both in the $EE$ and in the $TE$
power spectra. The latter has a much larger amplitude (because the temperature
fluctuations are much larger than the polarization fluctuations), allowing an
early detection by WMAP. However, it is affected by a much larger cosmic
variance (arising from the temperature term) and has an intrinsically weaker
dependence on $\tau$ ($TE\propto \tau$, $EE\propto \tau^2$); also, there is
only a partial correlation between $T$ and $E$
\citep{PlanckCollaborationXLVII2016}.

As a consequence, the low-$\ell$ $EE$ power spectrum dominates the constraints
on $\tau$.  The latest estimate is $\tau = 0.055\pm 0.009$
\citep{PlanckCollaborationXLVI2016}. This largely removes the tension with
constraints on $\tau$ derived from optical/UV data, implied by earlier
estimates yielding higher values of $\tau$
\citep{PlanckCollaborationXLVII2016}.

\subsection{$B$-mode from gravitational lensing}

The weak gravitational lensing of the CMB due to the intervening matter
distribution converts $E$-modes to the $B$-modes (also generating non-zero $TB$
and $EB$-correlations), in addition to smoothing the acoustic oscillations of
the power spectra of temperature and $E$-mode anisotropies and of adding power
at $\ell \simgt 3000$. This signal is totally independent from the existence of
primordial $B$ modes, i.e. of tensor modes in the early universe (see
Sect.~\ref{sect:Bmode}).

That due to gravitational lensing is the only $B$-mode signal detected so far.
The signal is weak and the observed signal is affected by noise, residual
foregrounds, systematics and cosmic variance. To ease the estimate of its power
spectrum a successful strategy consists in cross-correlating the total observed
$B$-mode map with a template constructed by combining a tracer of the
gravitational potential and an estimate of the primordial E-modes.
Cross-correlating with the cosmic infrared background (CIB), the SPTpol team
\citep{Hanson2013} reported the first estimate of the lensing B-mode power
spectrum. A similar cross-correlation result was obtained by the POLARBEAR
\citep{Ade2014a} and the ACTpol \citep{vanEngelen2015} groups.

The first direct evidence for polarization $B$-mode based on purely CMB
information was reported by the POLARBEAR collaboration \citep{Ade2014b}, using
the four-point correlations of $E$- and $B$-mode polarization. An improved
measurement from a blind analysis of data from the first two seasons of
POLARBEAR observations was presented by \citet{PolarbearCollaboration2017}.

\citet{PlanckCollaborationXV2016} detected lensing $B$-modes in the
\textit{Planck} data at a significance of $10\,\sigma$, using both a
cross-correlation with the CIB as a tracer of the lensing potential, as well as
a CMB-only approach  using the $TTEB$ trispectrum. This paper also presented a
measurement of the CMB lensing potential, significant at the $40\,\sigma$
level, using temperature and polarization data from the \textit{Planck} 2015
full-mission release.

An important outcome of the determination of the lensing potential are tight
constraints on the effective number of neutrino species, $N_{\rm eff}$, and on
the sum of neutrino masses, $\sum m_\nu$ \citep[for an exhaustive discussion
see][]{Abazajian2015}. Combining \textit{Planck} observations with other
astrophysical data \citet{PlanckCollaborationXIII2016}  find $N_{\rm
eff}=3.15\pm 0.23$ and $\sum m_\nu < 0.23\,$eV.

The constraints on $\sum m_\nu$ imply that neutrino masses have a very weak
effect on primordial CMB temperature anisotropies. Improvements on these
constraints will therefore be driven primarily by accurate measurements of
lensing $B$-modes. Next generation CMB experiments combined with improved
measurements of large scale structure will have the power to detect properties
of neutrinos with high accuracy, complementing the results of large laboratory
experiments.


\citet{PlanckCollaborationXLI2016} have produced a nearly all-sky template map
of the lensing-induced $B$-modes. This map was built combining two sets of
\textit{Planck} results: the measurements of the polarization $E$-modes and the
integrated mass distribution obtained via the reconstruction of the CMB lensing
potential.  It will be particularly useful for experiments searching for
primordial $B$-modes (see Sect.~\ref{sect:next}),  since it will allow an
estimate of the lensing-induced contribution to the measured total CMB
$B$-modes.

\begin{figure}
\vskip-0.3cm
\includegraphics[width=\textwidth]{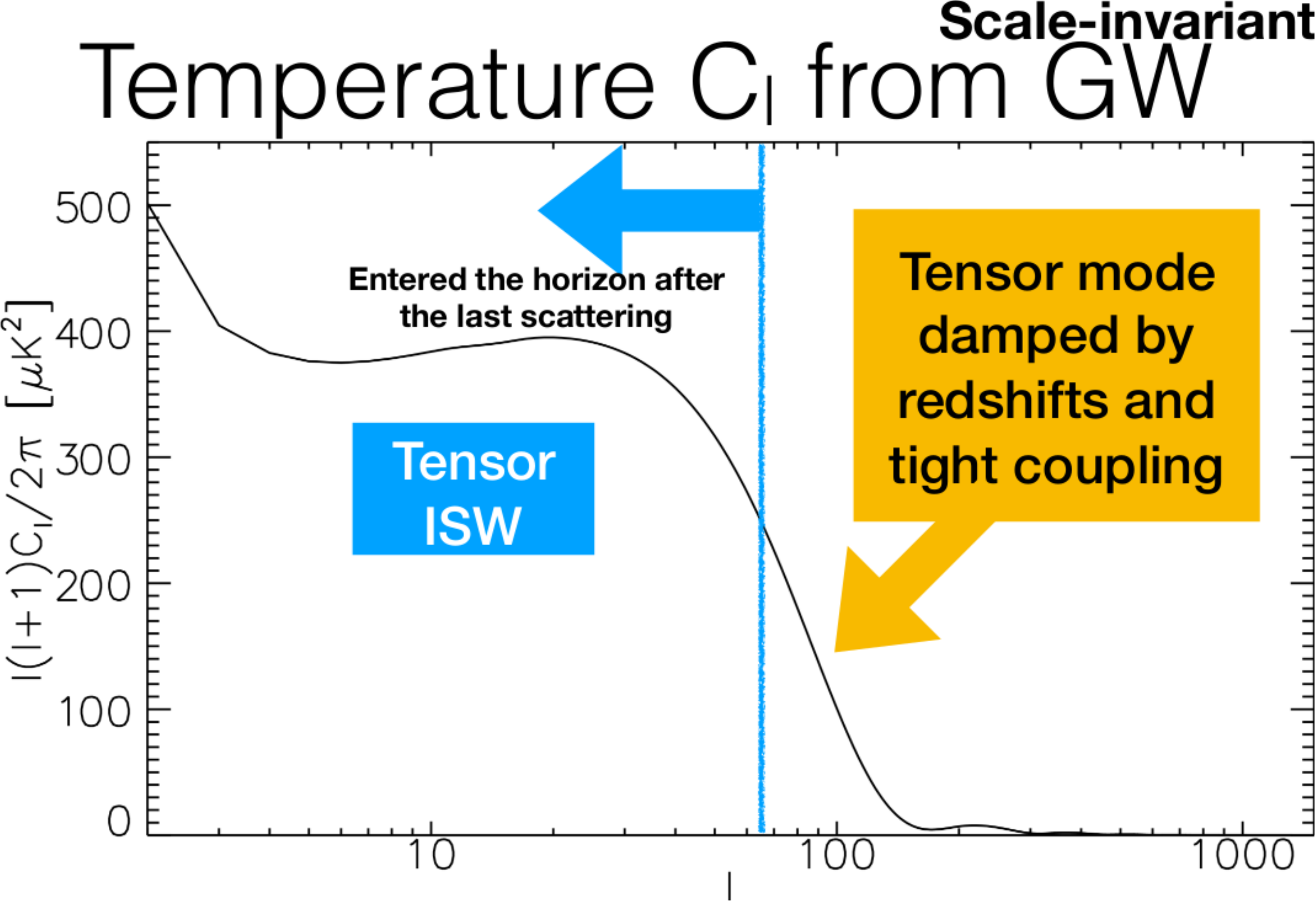}
\caption{Power spectrum of temperature anisotropies due to gravitational waves. Courtesy of Eiichiro
Komatsu; see his lectures given at the XIII School of Cosmology
November 12--18, 2017, Carg\`ese.}
\label{fig:GW}
\end{figure}

\subsection{Primordial $B$-modes}\label{sect:Bmode}

The simplest and canonical model for inflation, namely single-field slow-roll
inflation, made  a number of predictions that have been confirmed by all
current cosmological data:  primordial perturbations are adiabatic; the
spectrum of primordial perturbations is very nearly, but not precisely, scale
invariant; the distribution of primordial perturbations is very nearly
Gaussian.

The model also predicts the existence of a stochastic background of
gravitational waves. Their detection would constitute a fairly definitive test
of the prevailing single-field slow-roll models of inflation
\citep{KamionkowskiKovetz2016}. Primordial CMB $B$-mode polarization is the
specific signature of such gravitational waves and, as such, provides direct
information on the physics of primordial inflation.

In fact, density (scalar perturbations) have polarization amplitudes that
change only parallel or perpendicular to the wave vector, i.e. carry only
$E$-mode polarization (apart from the secondary effect of gravitational
lensing). But gravitational waves stretch the space, creating quadrupole
temperature anisotropy without velocity potential. The stretching in one
direction is accompanied by a contraction in the perpendicular direction. Since
the wavelength of light is also stretched, stretching/contraction results in a
drop or raise of the temperature. The polarization is parallel to the hot
regions and has components both parallel/perpendicular to the wave-number
vector ($E$-mode) and tilted by $45^\circ$ from that vector ($B$-mode). The two
components have similar amplitude, but on small scales $B$ is smaller than $E$
because $B$ vanishes on the horizon (see the lectures by E. Komatsu mentioned
above).

The amplitude of inflationary tensor modes is typically expressed in terms of
the tensor-to-scalar ratio $r =A_t/A_s$. The parameter $r$ provides a measure
of the expansion rate during inflation \citep{Abazajian2016}
\begin{equation}
H_{\rm infl} = 2.3\times 10^{13}\left(\frac{r}{0.01}\right)^{1/2}\,\hbox{GeV},
\end{equation}
which can be related to the energy scale, $V$, of inflation,
\begin{equation}
V = 1.04\times 10^{16}\left(\frac{r}{0.01}\right)^{1/4}\,\hbox{GeV}.
\end{equation}
The observation of primordial tensor modes would therefore associate inflation
with physics at the Grand Unified Theory (GUT) scale, estimated to be around
$10^{16}$\,GeV.

The anisotropy power spectra due to tensor perturbations are induced by
gravity, that is the only agent on super-horizon scales. When a perturbation
enters the horizon, the cosmological expansion damps the amplitude of tensor
modes via redshift. The tight coupling between electrons and photons (before
recombination) also damps temperature anisotropies. Those due to tensor
perturbations are not restored because gravitational waves are very weakly
coupled to photons. So only anisotropies on super-horizon scales at
recombination ($\ell \simgt 60$) survive (see Fig.~\ref{fig:GW}).

This limits the power of temperature anisotropies to constrain gravitational
waves since the sampling variance of the dominant scalar perturbations is large
at low $\ell$. Fortunately, CMB polarization provides an alternative route to
detecting the effect of gravitational waves on the CMB which is not limited by
cosmic variance of scalar perturbations since, as mentioned above, density
perturbations do not produce $B$-modes and those due to gravitational lensing
decline quickly at low multipoles.

There is no definitive prediction for the magnitude of $r$. However, some
arguments suggest $r \simgt 0.001$ \citep{KamionkowskiKovetz2016}. Models
currently of special interest (e.g., Starobinsky's $R^2$ and Higgs inflation)
predict $r \sim 0.003$. The detection of a signal at this level requires
extreme sensitivity, control of systematic effects and foreground removal.

\citet{BICEP2Collaboration2014} reported a significant ($> 5\,\sigma$) excess
of $B$-mode power over the $r=0$ lensed-$\Lambda$CDM expectation over the
multipole range $30< \ell < 150$. Some ($\simeq 1.7\,\sigma$) evidence against
the possibility that the signal can be accounted for by Galactic dust emission
was presented, based on the cross-spectrum against 100\,GHz maps from the
previous BICEP1 experiment. It was also pointed out that the detected $B$-mode
level was in excess of that expected from several dust models.

However \textit{Planck} observations at high Galactic latitude and, in
particular in a field centered on the BICEP2 region, found a level of polarized
dust emission at $353\,$GHz sufficient to account for the $150\,$GHz excess
observed by BICEP2, although with relatively low signal to noise
\citep{PlanckIntXXX2016}. A joint analysis of BICEP2/Keck and \textit{Planck}
data \citep{BICEP2KeckCollaborationPlanckCollaboration2015} showed that the
BICEP2/Keck $150\,$GHz polarization map was correlated with the \textit{Planck}
$353\,$GHz map of polarized dust emission, implying that the entire BICEP2
$B$-mode excess could be attributed to dust. This left a 95\% confidence upper
limit $r< 0.12$. An improved analysis of BICEP2 and Keck Array data, including
the 95\,GHz band \citep{BICEP2Keck2016}, has tightened the 95\% upper limit to
$r< 0.09$. Combining the BICEP2/Keck data with constraints from the
\textit{Planck} analysis of CMB temperature plus baryon acoustic oscillations
and other data yielded a combined limit $r <0.07$, again at the 95\% confidence
level; this limit is however somewhat model dependent.

\section{Next generation CMB experiments}\label{sect:next}

Several projects for next generation CMB experiments are being planned. The
main goal are accurate measurements of polarization anisotropies, with emphasis
on the search for primordial $B$-modes. However there is a renewed interest
also on new measurements of the CMB spectrum. A short summary of major projects
follows.

\subsection{LiteBIRD}

The \textbf{Lite} (Light) satellite for the studies of \textbf{B}-mode
polarization and \textbf{I}nflation from cosmic background \textbf{R}adiation
\textbf{D}etection\footnote{\url{http://litebird.jp/}} is, since September
2016, in the Japan Aerospace Exploration Agency/Institute of Space and
Astronautical Science (JAXA/ISAS) conceptual design phase, called ISAS
Phase-A1.

In February 2017 it was selected as one of 28 highest-priority large projects
by the Science Council of Japan and in July of the same year as one of 7
projects that should be listed in the ``Roadmap 2017 on promotion of large
research projects'' by the Ministry of Education, Culture, Sports, Science \&
Technology in Japan.

The mission goal is the verification of the inflation scenario by detecting the
primordial $B$-modes. More precisely, the mission requirements are:

\begin{itemize}

\item Measurement of the $B$-mode polarization power spectrum on large
    angular scales ($2\le \ell \le 200$).

\item Measurement of the tensor to scalar ratio $r$ with a precision
    $\sigma_r< 0.001$, without subtracting the gravitational lensing $B$
    modes.

\end{itemize}

The mission specifications are\footnote{See
\url{http://litebird.jp/wp-content/uploads/2012/03/LNPC_LiteBIRD_uozumi_small.pdf}
and \citet{Suzuki2018}.}:

\begin{itemize}

\item Operation at the second Lagrangian point of the Earth--Moon system
    (L2), located at about 1.5 million km from the Earth, directly `behind'
    it, as viewed from the Sun.

\item Full sky scan.

\item Optimized for large-angle measurements of the primordial $B$-modes.
    Telescope aperture size: 40\,cm; beam size $< 1^\circ$ over the full
    observing frequency range; field of view $10^\circ\times 20^\circ$.

\item High detector sensitivity: $3\,\mu$K$\cdot$arcmin, with margin.

\item Broad frequency range (40--400\,GHz, with 15 frequency bands) to
    characterize and remove polarized foreground emission.

\item Launch around 2025--2027.

\item Three years of operation.

\end{itemize}

\subsection{CORE}

The \textbf{C}osmic \textbf{OR}igin \textbf{E}xplorer (Principal Investigator:
J. Delabrouille; co-leads: P. de Bernardis and F.\,R.Bouchet) was proposed to
ESA in answer to the ``M5'' call for a medium-sized mission. Although it was
not selected, work on this project is still ongoing in view of a resubmission
at the next call.

The Executive Board includes representatives of many European countries (in
alphabetical order: Austria, Belgium, Denmark, Finland, France, Gemany,
Ireland, Italy, Netherlands, Norway, Poland, Portugal, Spain, Switzerland,
United Kingdom) as well as of the USA.

The baseline design specifications of CORE are \citep{Delabrouille2017}:

\begin{itemize}

\item A 1.2\,m telescope.

\item The instrument works at the diffraction limit, with angular resolutions
    ranging from $\simeq 2^\prime$ at the highest frequencies to $18^\prime$
    at the lowest. Note that \textit{Planck} did not work at the diffraction
    limit at its highest frequencies. This is why CORE reaches higher angular
    resolutions with a somewhat smaller telescope.

\item The observations are made with a single integrated focal-plane
    instrument, consisting of an array of 2100 cryogenically-cooled,
    linearly-polarised detectors. The full array aggregate CMB sensitivity is
    about $1.7\,\mu$K$\cdot$arcmin, 25 times better than \textit{Planck}.

\item There are 19 frequency channels, distributed over a broad frequency
    range, from 60 to 600\,GHz.  Frequency channels are chosen to cover a
    frequency range sufficient to  disentangle the  CMB from astrophysical
    foreground emission.  This is essential because even at frequencies where
    foreground emission is the lowest relative to the CMB, to reach the
    wanted sensitivity to the CMB polarization, more than 99\% of the
    Galactic emission must be removed from the observed maps, and/or the
    foreground emission contribution to the angular power spectrum of the
    observations must be modelled with $10^{-4}$ precision. Six frequency
    channels ranging from 130 GHz to 220 GHz are dedicated primarily to
    observing the CMB. Six channels from 60 to 115\,GHz monitor low-frequency
    foregrounds (polarised synchrotron, but also free-free and spinning dust
    in intensity, and in polarization if required). Seven channels ranging
    from 255 to 600\,GHz serve to monitor dust emission, and to map the
    cosmic infrared background (CIB) anisotropies that can serve as a tracer
    of mass for ``de-lensing'' CMB polarization $B$-modes.

\item The satellite will observe the sky for 4 years from a large Lissajous
    orbit around the L2 point. The scanning strategy combines three rotations
    of the spacecraft over different timescales. In this way about 50\% of
    the sky will be covered every few days.
\end{itemize}

\noindent Thanks to its substantially improved sensitivity and angular
resolution compared to \textit{Planck}, CORE will also provide a lot of
interesting information on Galactic emissions, on extragalactic radio sources
and dusty galaxies \citep{DeZotti2016} and on galaxy clusters detected via the
Sunyaev-Zeldovich effect \citep{Melin2017}.

\subsection{PICO}

The \textbf{P}robe of \textbf{I}nflation and \textbf{C}osmic \textbf{O}rigins
(formerly, CMB-Probe; P.I.: S. Hanany)\footnote{See
\url{https://zzz.physics.umn.edu/groups/ipsig/cmbprobe2016proposal} and
Presentations/Posters at
\url{https://zzz.physics.umn.edu/ipsig/start?&#presentationsposters}. } is one
of the 8 Probe-Scale (\$\,400M -- \$\,1000\,M) space missions whose study is
being funded by NASA. It will conduct a millimeter/sub-millimeter wave
polarimetic survey of the entire scale with:

\begin{itemize}

\item A 1.4\,m telescope.

\item 21 bands, with 25\% bandwidth, covering the frequency range from 21 to
    799\,GHz.

\item 12,356 transition edge sensor bolometers plus multiplexed readouts.

\end{itemize}

\noindent The instrument will work at the diffraction limit, with angular
resolutions ranging from $\simeq 38^\prime$ at the lowest frequency to $\simeq
1^\prime$ at the highest. The sensitivity to polarization in the central
(``CMB'') channels, from $\simeq 80$ to $\simeq 135\,$GHz is $\le
1.7\,\mu$K$\cdot$arcmin. The survey will be carried out for 4 years, from L2.

Like CORE, PICO will provide a rich harvest of new data also on astrophysical
foregrounds. Its $\simeq 1^\prime$ angular resolution at the highest
frequencies is ideal to detect the sub-mm emission of proto-clusters of dusty
galaxies. This will allow the investigation of early phases of cluster
evolution, before the establishment of the hot intergalactic medium that makes
them visible in X-rays or via the Sunyaev-Zeldovich effect.

\subsection{CMB-S4}

The CMB-Stage~IV \citep[CMB-S4;][]{Abazajian2016, Abitbol2017} is a US-led
ground based program building on Stage\,2 and 3 projects. It is expected to
start operating in $\sim 2021$. Like the projects mentioned above, it aims at
investigating primordial inflation by mapping the polarization of the CMB to
nearly the cosmic variance limit for a broad range of angular scales.

It targets to deploy $\sim 500,000$ effectively background-limited detectors,
spanning the 30 to 300\,GHz frequency range. Since any current single telescope
design cannot admit such a large number of detectors, CMB-S4 will use an array
of multiple telescopes. A conservative design of CMB-S4 would include both
small and large telescopes. Small telescopes have the role of setting the most
sensitive constraints on the degree scale, i.e. on the recombination peak of
polarization power spectra. To this end, they can be built with entirely
cryogenic optics, reducing detector noise due to optical loading from the
telescope.

Large telescope, will have primary apertures in the 2--10\,m diameter range, in
order to achieve an optical beam size in the range of 1--4 arcminutes at
$\simeq 100-150\,$GHz, This will allow the measurement of CMB power spectra up
$\ell_{\rm max}\sim 5000$, to meet many of the science requirements including
those that exploit gravitational lensing, measurements of the damping tail, and
galaxy cluster measurements. In particular, the arcmin angular resolution will
be optimally suited for ``de-lensing'', i.e. to measure and remove the
contribution to the $B$-mode power spectrum from gravitational lensing of CMB
$E$-modes.

The polarization sensitivity will be of $\sim 1\,\mu$K$\cdot$arcmin over
$\simgt 70\%$ of the sky, and better in deep fields. It is foreseen that,
thanks to these performances, CMB-S4 will reach uncertainties on the
tensor-to-scalar ratio $r$, on the effective number of neutrino species and on
the sum of neutrino masses of $\sigma(r)=0.001$, $\sigma(N_{\rm
eff})=0.02$--0.03 and $\sigma(\sum m_\nu)=20$--30\,meV, respectively.

The arcmin resolution is also crucial for extragalactic science, e.g. for
detecting high-$z$ strongly lensed galaxies, galaxy clusters via the
Sunyaev-Zeldovich effect and galaxy proto-clusters via the mm/sub-mm emission
of member galaxies. Thanks to its sensitivity and to its better resolution at
mm wavelengths, CMB-S4 will detect highest redshift sources than CORE or PICO,
up to $z\simgt 6$, as demonstrated by the SPT results \citep{Marrone2017}.

\subsection{PIXIE}

The \textbf{P}rimordial \textbf{I}nfla\textbf{XI}on (Inflation)
\textbf{E}xplorer \citep{Kogut2016} is conceptually different from all other
CMB polarization missions. It is designed to measure not only polarization, but
also the absolute spectrum of CMB emission, similarly to COBE-FIRAS, using a
reference blackbody to which the absolute emission of the sky is compared by
means of a Fourier transform spectrometer.

PIXIE covers the range from 30 to 6000\,GHz in 400 frequency bands. With an
overall sensitivity of about $5\,\mu$K$\cdot$arcmin, PIXIE is the least
sensitive of the CMB polarization space missions considered here (about 3 times
less than CORE, which directly translates into a sensitivity to the B-mode
power spectrum about 9 times worse if only noise is considered). This leaves
open the possibility of detecting primordial $B$-modes at the level $r\sim
0.01$ with PIXIE alone.

PIXIE will also measure the CMB $E$-mode on the largest scales, hence can
accurately determine the re-ionization optical depth $\tau$, but its angular
resolution of about $2^\circ.6$ is not sufficient for a clear observation of
the recombination peak of inflationary $B$-modes, nor for any lensing or
small-scale CMB $E$-mode science.

However, it will improve spectacularly over the COBE/FIRAS measurement of the
sky absolute brightness, allowing the detection of distortions of the CMB
spectrum orders of magnitude weaker than the COBE/FIRAS upper limits. The
detection of spectral distortions would have profound implications for our
understanding of physical processes taking place over a vast window in the
cosmological history \citep[e.g.,][]{DeZotti2016distortions, Chluba2016}.

The frequency range covered by PIXIE, extending to much higher frequencies than
any other CMB experiment, carries essential information about foreground
emission that cannot be obtained in any other way. Particularly valuable is the
accurate measurement of the CIB spectrum which is a measure of the overall
energy released by dust-obscured star formation and AGN accretion, and is
currently known with a $\sim 30\%$ uncertainty.

\subsection{Ground-based versus space-borne CMB experiments}

Large ground-based CMB programmes, like CMB-S4, can target sensitivities and
angular resolutions substantially better than space-borne missions: they can
operate over much longer times and use much larger telescopes.

On the other hand, they can necessarily cover a limited number of frequency
bands (the main atmospheric windows are centered around minima of atmospheric
emission at about 30, 90, 150, and 220 GHz) and this limits their capability of
removing foreground emissions. This is a crucial point: to have foreground
residuals below noise and/or cosmic variance uncertainties in bins of
$\Delta\ell \simeq 30\%$, foreground contamination must be reduced by at least
3 orders of magnitude in amplitude at $\ell \sim 10$, by 2 orders of magnitude
at $\ell \sim 100$ and 1 order of magnitude at $\ell \sim 1000$
\citep{Delabrouille2017}.

This is unlikely to be doable with ground-based experiments, which must thus
exploit only the cleanest sky regions, i.e. only a limited fraction of the sky.
This implies a higher sampling variance that, given the sensitivity of the
planned experiments, dominates for $\ell <2500$ and for $\ell <1000$ for $E$-
and $B$-modes, respectively.

Also, ground-based experiments are more liable to systematic effects. Space
missions completely avoid the complexity of atmospheric absorption, emission,
and fluctuations, minimise side-lobe pickup of emissions from the Earth, Sun,
and Moon and fluctuations of parts of the instrument that are optically coupled
to the detectors.

The history of CMB research has shown that ground-based and balloon-borne
observations are essential to build technological roadmaps and for observing
the small scales that are too costly from space. However, all the major steps
forward have been achieved by space missions: COBE, which confirmed the
blackbody spectrum of the CMB, ruling-out alternatives to the hot Big-Bang
scenario, and detected the first temperature anisotropies; WMAP which set the
stage for precision cosmology; \textit{Planck} which extracted essentially all
the cosmological information available in the CMB temperature spectrum on
scales $\simgt 5^\prime$.

\section{Conclusions}\label{sect:conclusions}

The recent advances in the detector technology make possible an increase in
sensitivity of CMB experiments by orders of magnitude. While \textit{Planck}
has already extracted the main cosmological information provided by CMB
temperature maps, the higher sensitivity is crucial for making substantial
progress on CMB polarization.

Next generation projects have the capability of measuring the CMB $E$-mode
polarization down to the cosmic variance limit over  a wide range of angular
scales.

Moreover, it will be possible to push the search for primordial $B$-modes down
to a tensor to scalar ratio $r \sim 0.001$. Primordial $B$-modes are the
current Holy Grail for cosmology and fundamental physics  because they provide
a measure of the energy scale driving the inflation, of order of $10^{16}\,$GeV
(far beyond the reach of accelerators on the ground: the maximum collision
energy reached by the Large Hadron Collider at CERN is $14\,\hbox{TeV}=
1.4\times 10^{14}\,$GeV) at $t\sim 10^{-35}\,$s.

Lensing $B$-modes may provide measurements of the neutrino absolute mass scale,
a determination of the neutrino mass hierarchy and strong constraints on
possible light relics like axions, sterile neutrinos and gravitinos.

\acknowledgements{I'm grateful to the the organizers of the Third Cosmology
School in Cracow for the kind invitation and the extraordinarily warm
hospitality. Thanks are due to Eiichiro Komatsu for having provided
Fig.~\ref{fig:GW}. Work supported in part by ASI/INAF agreement n.~2014-024-R.1
for the {\it Planck} LFI Activity of Phase E2 and by the ASI/Physics Department
of the university of Roma--Tor Vergata agreement n. 2016-24-H.0 for study
activities of the Italian cosmology community.}

\bibliographystyle{ptapap}
\bibliography{ptapapdoc}

\begin{thebibliography}{53}
\providecommand{\natexlab}[1]{#1}
\providecommand{\url}[1]{\texttt{#1}}
\providecommand{\urlprefix}{URL }
\providecommand{\eprint}[2][]{\url{#2}}

\bibitem[{{Abazajian} et~al.(2015)}]{Abazajian2015}
{Abazajian}, K.~N., et~al., \emph{{Inflation physics from the cosmic microwave
  background and large scale structure}}, \emph{Astroparticle Physics}
  \textbf{63}, 55 (2015), \eprint{1309.5381}

\bibitem[{{Abazajian} et~al.(2016)}]{Abazajian2016}
{Abazajian}, K.~N., et~al., \emph{{CMB-S4 Science Book, First Edition}},
  \emph{ArXiv e-prints}  (2016), \eprint{1610.02743}

\bibitem[{{Abitbol} et~al.(2017)}]{Abitbol2017}
{Abitbol}, M.~H., et~al., \emph{{CMB-S4 Technology Book, First Edition}},
  \emph{ArXiv e-prints}  (2017), \eprint{1706.02464}

\bibitem[{{Ade} et~al.(2014{\natexlab{a}})}]{Ade2014a}
{Ade}, P.~A.~R., et~al., \emph{{Evidence for Gravitational Lensing of the
  Cosmic Microwave Background Polarization from Cross-Correlation with the
  Cosmic Infrared Background}}, \emph{Physical Review Letters} \textbf{112},
  13, 131302 (2014{\natexlab{a}}), \eprint{1312.6645}

\bibitem[{{Ade} et~al.(2014{\natexlab{b}})}]{Ade2014b}
{Ade}, P.~A.~R., et~al., \emph{{Measurement of the Cosmic Microwave Background
  Polarization Lensing Power Spectrum with the POLARBEAR Experiment}},
  \emph{Physical Review Letters} \textbf{113}, 2, 021301 (2014{\natexlab{b}}),
  \eprint{1312.6646}

\bibitem[{{Barkats} et~al.(2014)}]{Barkats2014}
{Barkats}, D., et~al., \emph{{Degree-scale Cosmic Microwave Background
  Polarization Measurements from Three Years of BICEP1 Data}}, \emph{\apj}
  \textbf{783}, 67 (2014), \eprint{1310.1422}

\bibitem[{{Bennett} et~al.(2013)}]{Bennett2013}
{Bennett}, C.~L., et~al., \emph{{Nine-year Wilkinson Microwave Anisotropy Probe
  (WMAP) Observations: Final Maps and Results}}, \emph{\apjs} \textbf{208}, 20
  (2013), \eprint{1212.5225}

\bibitem[{{BICEP2 Collaboration}(2014)}]{BICEP2Collaboration2014}
{BICEP2 Collaboration}, \emph{{Detection of B-Mode Polarization at Degree
  Angular Scales by BICEP2}}, \emph{Physical Review Letters} \textbf{112}, 24,
  241101 (2014), \eprint{1403.3985}

\bibitem[{{BICEP2 Collaboration} \& {Keck Array
  Collaboration}(2016)}]{BICEP2Keck2016}
{BICEP2 Collaboration}, {Keck Array Collaboration}, \emph{{Improved Constraints
  on Cosmology and Foregrounds from BICEP2 and Keck Array Cosmic Microwave
  Background Data with Inclusion of 95 GHz Band}}, \emph{Physical Review
  Letters} \textbf{116}, 3, 031302 (2016), \eprint{1510.09217}

\bibitem[{{BICEP2/Keck Collaboration and Planck
  Collaboration}(2015)}]{BICEP2KeckCollaborationPlanckCollaboration2015}
{BICEP2/Keck Collaboration and Planck Collaboration}, \emph{{Joint Analysis of
  BICEP2/Keck Array and Planck Data}}, \emph{Physical Review Letters}
  \textbf{114}, 10, 101301 (2015), \eprint{1502.00612}

\bibitem[{{Chluba}(2016)}]{Chluba2016}
{Chluba}, J., \emph{{Which spectral distortions does {$\Lambda$}CDM actually
  predict?}}, \emph{\mnras} \textbf{460}, 227 (2016), \eprint{1603.02496}

\bibitem[{{CORE Collaboration}(2015)}]{Finelli2016}
{CORE Collaboration}, \emph{{A Measurement of Secondary Cosmic Microwave
  Background Anisotropies from the 2500 Square-degree SPT-SZ Survey}},
  \emph{\apj} \textbf{799}, 177 (2015), \eprint{1408.3161}

\bibitem[{{Crites} et~al.(2015)}]{Crites2015}
{Crites}, A.~T., et~al., \emph{{Measurements of E-Mode Polarization and
  Temperature-E-Mode Correlation in the Cosmic Microwave Background from 100
  Square Degrees of SPTpol Data}}, \emph{\apj} \textbf{805}, 36 (2015),
  \eprint{1411.1042}

\bibitem[{{De Zotti} et~al.(2015)}]{DeZotti2015}
{De Zotti}, G., et~al., \emph{{Extragalactic sources in Cosmic Microwave
  Background maps}}, \emph{\jcap} \textbf{6}, 018 (2015), \eprint{1501.02170}

\bibitem[{{De Zotti} et~al.(2016{\natexlab{a}})}]{DeZotti2016distortions}
{De Zotti}, G., et~al., \emph{{Another look at distortions of the Cosmic
  Microwave Background spectrum}}, \emph{\jcap} \textbf{3}, 047
  (2016{\natexlab{a}}), \eprint{1512.04816}

\bibitem[{{De Zotti} et~al.(2016{\natexlab{b}})}]{DeZotti2016}
{De Zotti}, G., et~al., \emph{{Exploring Cosmic Origins with CORE:
  Extragalactic sources in Cosmic Microwave Background maps}}, \emph{ArXiv
  e-prints}  (2016{\natexlab{b}}), \eprint{1609.07263}

\bibitem[{{Delabrouille} et~al.(2017)}]{Delabrouille2017}
{Delabrouille}, J., et~al., \emph{{Exploring Cosmic Origins with CORE: Survey
  requirements and mission design}}, \emph{ArXiv e-prints}  (2017),
  \eprint{1706.04516}

\bibitem[{{George} et~al.(2015)}]{George2015}
{George}, E.~M., et~al., \emph{{A Measurement of Secondary Cosmic Microwave
  Background Anisotropies from the 2500 Square-degree SPT-SZ Survey}},
  \emph{\apj} \textbf{799}, 177 (2015), \eprint{1408.3161}

\bibitem[{{Hanson} et~al.(2013)}]{Hanson2013}
{Hanson}, D., et~al., \emph{{Detection of B-Mode Polarization in the Cosmic
  Microwave Background with Data from the South Pole Telescope}},
  \emph{Physical Review Letters} \textbf{111}, 14, 141301 (2013),
  \eprint{1307.5830}

\bibitem[{{Henning} et~al.(2017)}]{Henning2017}
{Henning}, J.~W., et~al., \emph{{Measurements of the Temperature and E-Mode
  Polarization of the CMB from 500 Square Degrees of SPTpol Data}}, \emph{ArXiv
  e-prints}  (2017), \eprint{1707.09353}

\bibitem[{{Hu} \& {White}(1997)}]{HuWhite1997}
{Hu}, W., {White}, M., \emph{{A CMB polarization primer}}, \emph{\na}
  \textbf{2}, 323 (1997), \eprint{astro-ph/9706147}

\bibitem[{{Kamionkowski} et~al.(1997){Kamionkowski}, {Kosowsky}, \&
  {Stebbins}}]{Kamionkowski1997}
{Kamionkowski}, M., {Kosowsky}, A., {Stebbins}, A., \emph{{Statistics of cosmic
  microwave background polarization}}, \emph{\prd} \textbf{55}, 7368 (1997),
  \eprint{astro-ph/9611125}

\bibitem[{{Kamionkowski} \& {Kovetz}(2016)}]{KamionkowskiKovetz2016}
{Kamionkowski}, M., {Kovetz}, E.~D., \emph{{The Quest for B Modes from
  Inflationary Gravitational Waves}}, \emph{\araa} \textbf{54}, 227 (2016),
  \eprint{1510.06042}

\bibitem[{{Knox}(1995)}]{Knox1995}
{Knox}, L., \emph{{Determination of inflationary observables by cosmic
  microwave background anisotropy experiments}}, \emph{\prd} \textbf{52}, 4307
  (1995), \eprint{astro-ph/9504054}

\bibitem[{{Kogut} et~al.(2016)}]{Kogut2016}
{Kogut}, A., et~al., \emph{{The Primordial Inflation Explorer (PIXIE)}}, in
  Society of Photo-Optical Instrumentation Engineers (SPIE) Conference Series,
  \emph{\procspie}, volume 9904, 99040W (2016)

\bibitem[{{Larson} et~al.(2011)}]{Larson2011}
{Larson}, D., et~al., \emph{{Seven-year Wilkinson Microwave Anisotropy Probe
  (WMAP) Observations: Power Spectra and WMAP-derived Parameters}},
  \emph{\apjs} \textbf{192}, 16 (2011), \eprint{1001.4635}

\bibitem[{{Liu} et~al.(2001)}]{Liu2001}
{Liu}, G.-C., et~al., \emph{{Polarization of the Cosmic Microwave Background
  from Nonuniform Reionization}}, \emph{\apj} \textbf{561}, 504 (2001),
  \eprint{astro-ph/0101368}

\bibitem[{{Louis} et~al.(2017)}]{Louis2017}
{Louis}, T., et~al., \emph{{The Atacama Cosmology Telescope: two-season ACTPol
  spectra and parameters}}, \emph{\jcap} \textbf{6}, 031 (2017),
  \eprint{1610.02360}

\bibitem[{{Marrone} et~al.(2017)}]{Marrone2017}
{Marrone}, D.~P., et~al., \emph{{Galaxy growth in a massive halo in the first
  billion years of cosmic history}}, \emph{ArXiv e-prints}  (2017),
  \eprint{1712.03020}

\bibitem[{{Melin} et~al.(2017)}]{Melin2017}
{Melin}, J.-B., et~al., \emph{{Exploring Cosmic Origins with CORE: Cluster
  Science}}, \emph{ArXiv e-prints}  (2017), \eprint{1703.10456}

\bibitem[{{Nolta} et~al.(2009)}]{Nolta2009}
{Nolta}, M.~R., et~al., \emph{{Five-Year Wilkinson Microwave Anisotropy Probe
  Observations: Angular Power Spectra}}, \emph{\apjs} \textbf{180}, 296 (2009),
  \eprint{0803.0593}

\bibitem[{{Page} et~al.(2007)}]{Page2007}
{Page}, L., et~al., \emph{{Three-Year Wilkinson Microwave Anisotropy Probe
  (WMAP) Observations: Polarization Analysis}}, \emph{\apjs} \textbf{170}, 335
  (2007), \eprint{astro-ph/0603450}

\bibitem[{{Planck Collaboration Int. XXX}(2016)}]{PlanckIntXXX2016}
{Planck Collaboration Int. XXX}, \emph{{Planck intermediate results. XXX. The
  angular power spectrum of polarized dust emission at intermediate and high
  Galactic latitudes}}, \emph{\aap} \textbf{586}, A133 (2016),
  \eprint{1409.5738}

\bibitem[{{Planck Collaboration XI}(2016)}]{PlanckCollaborationXI2016}
{Planck Collaboration XI}, \emph{{Planck 2015 results. XI. CMB power spectra,
  likelihoods, and robustness of parameters}}, \emph{\aap} \textbf{594}, A11
  (2016), \eprint{1507.02704}

\bibitem[{{Planck Collaboration XIII}(2016)}]{PlanckCollaborationXIII2016}
{Planck Collaboration XIII}, \emph{{Planck 2015 results. XIII. Cosmological
  parameters}}, \emph{\aap} \textbf{594}, A13 (2016), \eprint{1502.01589}

\bibitem[{{Planck Collaboration XLI}(2016)}]{PlanckCollaborationXLI2016}
{Planck Collaboration XLI}, \emph{{Planck intermediate results. XLI. A map of
  lensing-induced B-modes}}, \emph{\aap} \textbf{596}, A102 (2016),
  \eprint{1512.02882}

\bibitem[{{Planck Collaboration XLVI}(2016)}]{PlanckCollaborationXLVI2016}
{Planck Collaboration XLVI}, \emph{{Planck intermediate results. XLVI.
  Reduction of large-scale systematic effects in HFI polarization maps and
  estimation of the reionization optical depth}}, \emph{\aap} \textbf{596},
  A107 (2016), \eprint{1605.02985}

\bibitem[{{Planck Collaboration XLVII}(2016)}]{PlanckCollaborationXLVII2016}
{Planck Collaboration XLVII}, \emph{{Planck intermediate results. XLVII. Planck
  constraints on reionization history}}, \emph{\aap} \textbf{596}, A108 (2016),
  \eprint{1605.03507}

\bibitem[{{Planck Collaboration XV}(2016)}]{PlanckCollaborationXV2016}
{Planck Collaboration XV}, \emph{{Planck 2015 results. XV. Gravitational
  lensing}}, \emph{\aap} \textbf{594}, A15 (2016), \eprint{1502.01591}

\bibitem[{{Planck Collaboration XVI}(2014)}]{PlanckCollaborationXVI2014}
{Planck Collaboration XVI}, \emph{{Planck 2013 results. XVI. Cosmological
  parameters}}, \emph{\aap} \textbf{571}, A16 (2014), \eprint{1303.5076}

\bibitem[{{Planck Collaboration XVII}(2016)}]{PlanckCollaborationXVII2016}
{Planck Collaboration XVII}, \emph{{Planck 2015 results. XVII. Constraints on
  primordial non-Gaussianity}}, \emph{\aap} \textbf{594}, A17 (2016),
  \eprint{1502.01592}

\bibitem[{{Planck Collaboration XX}(2016)}]{PlanckCollaborationXX2016}
{Planck Collaboration XX}, \emph{{Planck 2015 results. XX. Constraints on
  inflation}}, \emph{\aap} \textbf{594}, A20 (2016), \eprint{1502.02114}

\bibitem[{{Planck Collaboration XXII}(2014)}]{PlanckCollaborationXXII2014}
{Planck Collaboration XXII}, \emph{{Planck 2013 results. XXII. Constraints on
  inflation}}, \emph{\aap} \textbf{571}, A22 (2014), \eprint{1303.5082}

\bibitem[{{Planck Collaboration XXIV}(2014)}]{PlanckCollaborationXXIV2014}
{Planck Collaboration XXIV}, \emph{{Planck 2013 results. XXIV. Constraints on
  primordial non-Gaussianity}}, \emph{\aap} \textbf{571}, A24 (2014),
  \eprint{1303.5084}

\bibitem[{{Polarbear Collaboration}(2014)}]{PolarbearCollaboration2014}
{Polarbear Collaboration}, \emph{{A Measurement of the Cosmic Microwave
  Background B-mode Polarization Power Spectrum at Sub-degree Scales with
  POLARBEAR}}, \emph{\apj} \textbf{794}, 171 (2014), \eprint{1403.2369}

\bibitem[{{Rybicki} \& {Lightman}(1979)}]{RybickiLightman1979}
{Rybicki}, G.~B., {Lightman}, A.~P., {Radiative processes in astrophysics}
  (1979)

\bibitem[{{Seljak}(1997)}]{Seljak1997}
{Seljak}, U., \emph{{Measuring Polarization in the Cosmic Microwave
  Background}}, \emph{\apj} \textbf{482}, 6 (1997), \eprint{astro-ph/9608131}

\bibitem[{{Smoot} et~al.(1992)}]{Smoot1992}
{Smoot}, G.~F., et~al., \emph{{Structure in the COBE differential microwave
  radiometer first-year maps}}, \emph{\apjl} \textbf{396}, L1 (1992)

\bibitem[{{Suzuki} et~al.(2018)}]{Suzuki2018}
{Suzuki}, A., et~al., \emph{{The LiteBIRD Satellite Mission - Sub-Kelvin
  Instrument}}, \emph{ArXiv e-prints}  (2018), \eprint{1801.06987}

\bibitem[{{The POLARBEAR Collaboration}(2017)}]{PolarbearCollaboration2017}
{The POLARBEAR Collaboration}, \emph{{A Measurement of the Cosmic Microwave
  Background B-mode Polarization Power Spectrum at Subdegree Scales from Two
  Years of polarbear Data}}, \emph{\apj} \textbf{848}, 121 (2017),
  \eprint{1705.02907}

\bibitem[{{van Engelen} et~al.(2015)}]{vanEngelen2015}
{van Engelen}, A., et~al., \emph{{The Atacama Cosmology Telescope: Lensing of
  CMB Temperature and Polarization Derived from Cosmic Infrared Background
  Cross-correlation}}, \emph{\apj} \textbf{808}, 7 (2015), \eprint{1412.0626}

\bibitem[{{White} et~al.(1994){White}, {Scott}, \& {Silk}}]{White1994}
{White}, M., {Scott}, D., {Silk}, J., \emph{{Anisotropies in the Cosmic
  Microwave Background}}, \emph{\araa} \textbf{32}, 319 (1994)

\bibitem[{{Zaldarriaga} \& {Seljak}(1997)}]{ZaldarriagaSeljak1997}
{Zaldarriaga}, M., {Seljak}, U., \emph{{All-sky analysis of polarization in the
  microwave background}}, \emph{\prd} \textbf{55}, 1830 (1997),
  \eprint{astro-ph/9609170}

\end{thebibliography}

\end{document}